\documentclass[aps,twocolumn,groupedaddress,superscriptaddress]{revtex4}

\usepackage{graphicx}

\begin{document}

\title{Negative refractive index response of weakly and strongly coupled optical metamaterials.}

\author{Jiangfeng Zhou}
\affiliation{Ames Laboratory and Department of Physics and
Astronomy,Iowa State University, Ames, Iowa 50011}

\author{Thomas Koschny}
\affiliation{Ames Laboratory and Department of Physics and
Astronomy,Iowa State University, Ames, Iowa 50011}
\affiliation{Department of Materials Science and Technology, University of Crete and Institute of Electronic Structure and Laser¨CFoundation for Research and Technology Hellas (FORTH), 71110 Heraklion, Crete, Greece}

\author{Maria Kafesaki}
\affiliation{Department of Materials Science and Technology, University of Crete and Institute of Electronic Structure and Laser¨CFoundation for Research and Technology Hellas (FORTH), 71110 Heraklion, Crete, Greece}

\author{Costas M. Soukoulis}
\affiliation{Ames Laboratory and Department of Physics and
Astronomy,Iowa State University, Ames, Iowa 50011}
\affiliation{Department of Materials Science and Technology, University of Crete and Institute of Electronic Structure and Laser¨CFoundation for Research and Technology Hellas (FORTH), 71110 Heraklion, Crete, Greece}

\date{\today}

\begin{abstract}
We present a detailed study of the retrieved optical parameters, electrical
permittivity, $\varepsilon$, magnetic permeability, $\mu$, and refractive
index, $n$, of the coupled fishnet metamaterial structures as a function of the
separation between layers.
For the weak coupling case, the retrieved parameters are very close to the
one-functional-layer results and converge relatively fast.
For the strong coupling case, the retrieved parameters are completely different
than the one unit fishnet results.
We also demonstrate that the high value of the figure of merit
($\mathrm{FOM}=|\mathrm{Re}(n)/\mathrm{Im}(n)|$) for the strongly coupled
structures is due to the fact that the real part of the negative $n$ moves away
from the maximum of the imaginary part of $n$ (close to the resonance), where
the losses are high.
\end{abstract}


\maketitle

\section{Introduction}
Metamaterials are artificially engineered structures that have properties, such
as negative refractive index, $n$, nonexistent in natural materials. The recent
development of metamaterials \cite{Adv_Mat_soukoulis_review_2006} with negative
$n$ confirms that structures can be fabricated and interpreted as having both a
negative permittivity, $\epsilon$, and a negative permeability, $\mu$,
simultaneously. Since the original microwave experiments for the demonstration
of negative index behavior in split ring resonators (SRRs) and wire structures,
new designs have been introduced, such as fishnet, that have pushed the
existence of the negative refraction at Thz and optical wavelengths
\cite{PRL_zhang_2005_137404,dolling_OL_2006,science_soukoulis_2006,
      Opl_Chettiar_2007,PRB_kafesaki_2007_235114,zhang_nature_2008}.
Most of the experiments with the fishnet structure measure transmission, $T$,
and reflection, $R$, and use the retrieval procedure
\cite{PRB_smith_retrieval_2002,PRB_koschny_retrieval_2005,PRE_Smith_retrieval_2005,PRE_Chen_retrieval_2004}
to obtain the effective parameters, $\epsilon$, $\mu$, and $n$.  Although,
stacking of three \cite{dolling_OL_2007_three_layer_LHM}, four
\cite{Liu_Nat_mat_2008}, 10 \cite{katsarakis_OL_2005} functional layers and
recently fabricated \cite{zhang_nature_2008} 10-functional layer fishnets (21
layers of silver and $\mathrm{MgF}_2$) have been realized, they do not
constitute a bulk metamaterial. Even the thickest fabricated
\cite{zhang_nature_2008} fishnet structure only has a total thickness, 830 nm,
half of the wavelength ($\lambda$=1700 nm). Here, we report a detailed study of
the weakly and strongly coupled fishnets to understand the origin of negative
$n$, as well the mechanism of low losses (that is, high figure of merit (FOM))
for the weakly and strongly coupled fishnets. We also study the convergence of
the retrieval parameter ($\epsilon$, $\mu$ and $n$) as the number of unit cells
(layers) increases. For the weakly coupled structures, the convergence results
for $n$ and FOM are close to the single unit cell. As expected, for the
strongly coupled structures, hybridization is observed and the retrieval
results for $n$ and FOM are completely different from the single unit cell. We
demonstrate that the high value of FOM for the strongly coupled structure is
due to the fact that the real part of negative $n$ moves away from the maximum
of the imaginary part of $n$ (close to the resonance), where the losses are
high.

The idea of left-handed materials, i.e., materials with both negative
$\epsilon$ and negative $\mu$, where the electric field ({\bf E}), magnetic
field ({\bf H}), and wave vector ({\bf k}) form a left-handed coordinate system
was developed by Veselago \cite{Veselago_1968} decades ago. However, it was
only recently that such materials were investigated experimentally at high
frequencies
\cite{PRL_zhang_2005_137404,dolling_OL_2006,science_soukoulis_2006,
      Opl_Chettiar_2007,PRB_kafesaki_2007_235114,zhang_nature_2008},
and the field is driven by a wide range of new applications, such as
ultrahigh-resolution imaging system \cite{zhang_nmat_supperlense_2008},
cloaking devices \cite{Pendry_science_cloak_2006,smith_science_cloak_2006}, and
quantum levitation \cite{leonhardt_quantum_levitation_2007}.  Realizing these
applications, several goals must be achieved: three-dimensional rather than
planar structure, isotropic design, and reduction of loss.

Most of the metamaterials exhibiting artificial magnetism
\cite{sci_yen_Zhang_Smith_2004,science_Linden_100THz,
OL_Dolling_wire_pair,katsarakis_OL_2005,Liu_Nat_mat_2008} and a negative
refractive index, $n$, at THz and optical frequencies
\cite{PRL_zhang_2005_137404,dolling_OL_2006,science_soukoulis_2006,Opl_Chettiar_2007,
OL_Dolling_wire_pair}, consist of only a functional layer. The number of actual
layers $M=2\times N+1$, where $N$ is the number of functional layers. The first
five-functional-layer of SRRs operating at 6 THz was published
\cite{katsarakis_OL_2005} in 2005, and four layers of SRRs operating at 70 THz
\cite{Liu_Nat_mat_2008} was published in 2008. The first three-functional-layer
of fishnets (7 layers of silver and $\mathrm{MgF}_2$) operating at 200 THz was
published \cite{dolling_OL_2007_three_layer_LHM} in 2007, and recently a
10-functional-layer of fishnets (21 layers of silver and $\mathrm{MgF}_2$)
operating at 200 THz was fabricated \cite{zhang_nature_2008}.  However, it is
very important to study how the optical properties ($\epsilon$, $\mu$ and $n$)
change as one increases the number of layers. How many layers are needed to
achieve convergence of the optical properties and one can call this
metamaterial bulk? How do optical properties behave as one changes the distance
between two neighboring fishnets? If the distance is small, we have a strong
coupling case, and one achieves the photonic crystal limits. The convergence of
optical properties is slow, and more importantly, it does not convergence to
the isolated fishnet case. What is the mechanism for negative $n$ in the strong
coupling limit?

In this paper, we present a detailed study of the retrieved optical
parameters, $\epsilon$, $\mu$, and $n$ of the single fishnet metamaterial
structures as a function of the size of the unit cell.  We find that as the
size of the unit cell decreases, the magnitude of the retrieved effective
parameters increases. In order to understand the underlying physics of the
coupled structures, we study the retrieved parameters of the coupled fishnets
as a function of the distance between them. Finally, we study the convergence
of the retrieved parameters as the number of the unit cell increases for the
weakly and strongly coupled structures. For the weakly coupling case, the
retrieved parameters are very close to the one-functional layer results and
converge relatively fast. For the strong coupling case, the retrieved
parameters are completely different than the one unit fishnet results. The
strong coupling case explains the recently observed negative refractive index
in the 21-layer fishnet structure \cite{zhang_nature_2008}, especially the high
FOM, due to the periodicity effects, as will be shown below.

\section{Weakly and strongly coupled fishnets}

\begin{figure}[htb]
 \centering
  \includegraphics[width=9cm]{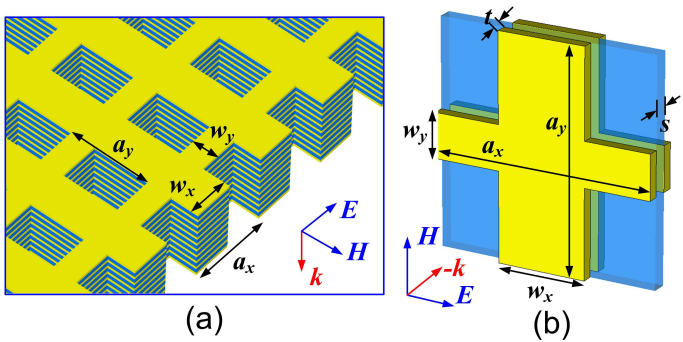}
  \caption{
    (a) Schematic of a fishnet structure with 11 metallic layers,
    (b) a single unit cell with geometric parameters marked on it.
  }
  \label{fig_geom}
\end{figure}

In Fig. \ref{fig_geom} we present a schematic graph of the unit-cell
of the fishnet structure. The size of the unit cell along the
propagation direction is $a_z$. $a_z$ is larger than the sum of the
thickness of the metallic and the dielectric layers $2t+s$, where
$t$ and $s$ are the thicknesses of the metal and dielectric layers,
respectively. Notice the propagation direction is perpendicular
to the plane of the fishnet.\\

In most of the experiments measuring the $T$ and $R$ of the fishnet
structure
\cite{PRL_zhang_2005_137404,dolling_OL_2006,science_soukoulis_2006,Opl_Chettiar_2007,PRB_kafesaki_2007_235114,
OL_Dolling_wire_pair}, there is only one layer of the sample
measured. In this case, the unit cell size along the propagation
direction, $a_z$, is undefined. We have shown
\cite{zhou_photon_nano_2008} that, as $a_z$ decreases, the magnitude
of the retrieved parameters increases. It is well known from
electronic systems that a monolayer of a  surface can exhibit
different properties from the bulk (many layers). So it is very
important to systematically study whether the optical parameters of
a single layer really correspond to the many layers system. We will
study the weak and strong coupling limit of the two-layer fishnet structure.\\

\begin{figure}[htb]
 \centering
  \includegraphics[width=8cm]{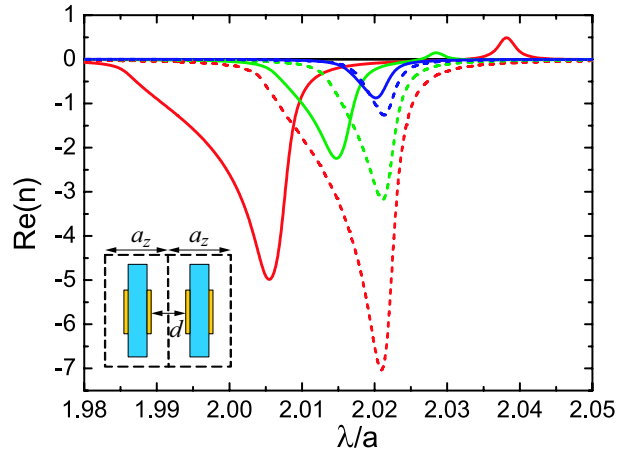}
  \caption{
    Retrieved real part of refractive index, $n$, from simulated data
    using unit cell size in the propagation direction $a_z=a/15$ (red),
    $a_z=2a/15$ (green) and $a_z=4a/15$ (blue). Both one layer (dashed)
    and two layers (solid) results are shown. The distances between two unit cells are
    $d=a_z-(2t+s)=$0.04a, 0.11a, and 0.24a, respectively.
    The other geometric
    parameters are given by $a_x=a_y=a$, $w_x=4a/15$,
    $w_y=3a/5$, $s=a/60$, $t=a/300$, and the dielectric constant of
    the spacer is $\epsilon_r=5$.
  }
  \label{fig2_n_1_2_cell}
\end{figure}

Figure \ref{fig2_n_1_2_cell} shows the real part of the effective
refractive index, Re($n$), as a function of $\lambda/a$, for one
layer and two layers of the fishnet structure described in Fig.
\ref{fig_geom}, with different distances between the unit cells.
Notice the normalized resonance wavelength $\lambda_m/a\approx2.02$,
i.e. wavelength with maximum $|\mathrm{Re}(n)|$, for one layer
shifts only slightly when the size of the unit cell increases, but
the magnitude of $|\mathrm{Re}(n)|$ decreases dramatically. For the
two layers, when the distance, $d$, between them is large
($d/a$=0.24, blue solid curve), the coupling between the two layers
is weak and, therefore, the refractive index, Re($n$), approaches
the one layer simulation results. When the distance between the two
layers becomes smaller ($d/a$=0.04, red solid curve) and the
coupling becomes stronger, hybridization takes place and two
resonance modes exist, one at $\lambda/2=2.005$, which gives
$\mathrm{Re}(n)<0$; and one at 2.040, which has $\mathrm{Re}(n)>0$.
The difference in value of the two resonance frequencies becomes
larger as the distance between them
decreases.\\
Another very important issue is how fast the optical retrieval
properties ($\epsilon$, $\mu$ and $n$) converge as the number of
unit cells increases. We will present results for two cases, one for
the weakly coupled fishnets.\\

The only design that gave negative $n$ at THz and optical
frequencies is the so-called ``double-fishnet" structure, which
consists of a pair of metal fishnets separated by a dielectric
spacer
\cite{PRL_zhang_2005_137404,dolling_OL_2006,science_soukoulis_2006,
      Opl_Chettiar_2007,PRB_kafesaki_2007_235114,zhang_nature_2008}.
For the incident polarization shown in Fig. \ref{fig_geom}, the thin
metallic wires along the x-axis, parallel to the incident electric
field, ${\bf E}$, excite the plasmonic response and produce negative
permittivity $\epsilon$ up to the plasma frequency. Negative $\mu$
is obtained from the wires along the y-axis, parallel to the
incident magnetic field $\bf{H}$. At the magnetic resonance
frequency, the two parallel bars sustain anti-parallel currents
(along x-axis), providing a magnetic field ${\bf B^\prime}$, mainly
between the plates and directly opposite to the external magnetic
field, {\bf H}. The electric field, because of the opposite charges
accumulate at the ends of the two metallic bars, is expected to be
confined within the space between the plates and near the end
points. Indeed, obtained simulations confirm this picture.\\

\begin{figure}[htb]
 \centering
  \includegraphics[width=8cm]{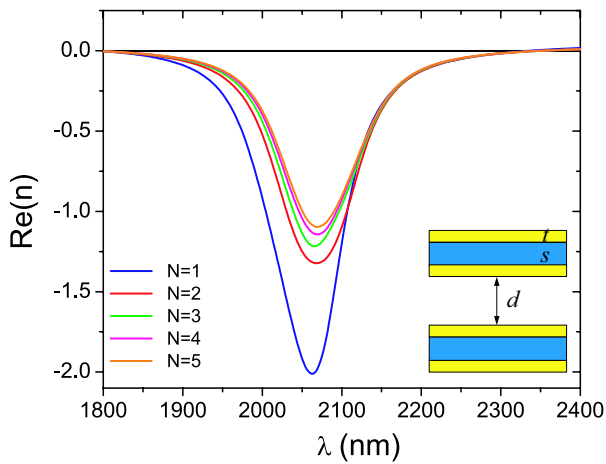}
  \caption{
    Retrieved real part of effective refractive index, $\mathrm{Re}(n)$
    for one layer (red solid), four layers (blue dashed),
    eight layers (green dotted) and ten layers (black dash-dotted)
    of the fishnet structure.
    The geometric parameters are $a_x=a_y=860$ nm, $w_x=565$ nm,
    $w_y=$ 265nm, $s=50$ nm, $t=30$ nm, $d=90$ nm, and the spacer is
    made from $\mathrm{MgF}_2$ with the dielectric constant
    $\epsilon_r=1.9$.
    The functional layers are separated by vacuum layers with
    thickness $d_0$ as shown in the inset.
  }
  \label{fig3_n_1_10_cell}
\end{figure}

In Fig. \ref{fig3_n_1_10_cell} we present the retrieved results for
the effective refractive index, $\mathrm{Re}(n)$, as a function of
$\lambda$ for different numbers of functional layers (N=1, 2, 3, 4 and 5)
for weakly coupled fishnets system. The parameters are exactly the same as the strongly coupled case, that will be discussed below, but the spacing between the functional layers is $d=90$ nm. As can be seen in Fig. \ref{fig3_n_1_10_cell}, the retrieved results for $\mathrm{Re}(n)$ converge very fast (N=2) and the convergence results agree with the results of the one functional layer of the fishnet.

\begin{figure}[htb]
 \centering
  \includegraphics[width=8cm]{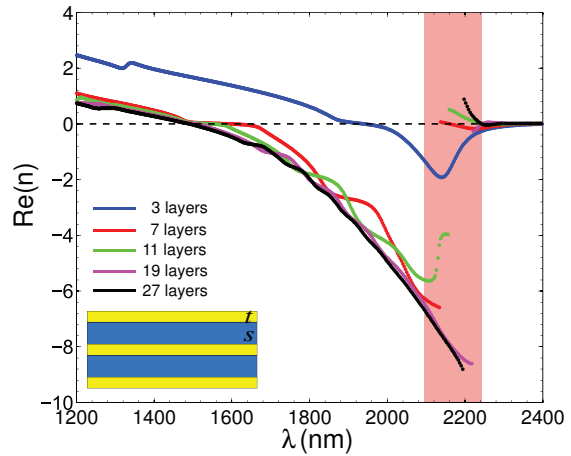}
  \caption{
    The retrieved real part of $n$ for 3, 7, 11, 19 and 27
    layers strongly coupled fishnet structure. The geometric
    parameters are $a_x=a_y=860$ nm, $w_x=565$ nm, $w_y=$ 265nm,
    $s=50$ nm and $t=30$ nm, and the spacer is made from
    $\mathrm{MgF}_2$ with the dielectric constant $\epsilon_r=1.9$.
    The shadow region shows where the discontinuity happens.
  }
  \label{fig_n_converge}
\end{figure}

When the fishnets strongly interact, it's not clear what the mechanism is for
giving negative $n$. As discussed in Fig.  \ref{fig2_n_1_2_cell}, the isolated
fishnet resonance frequency hybridizes into two different modes. The
antisymmetric mode gives weak resonance with $n\approx 0$, while the symmetric
mode gives a strong resonance with a strong negative $n$. In Fig.
\ref{fig_n_converge} we present results for the retrieved, $\mathrm{Re}(n)$,
for different number of layers (3 to 27) for the recently fabricated
\cite{zhang_nature_2008} negative index structure. Notice in the low wavelength
limit (between 1200 to 2100 nm), convergence of $n$ is obtained and agrees with
experimental results of Ref.  \cite{zhang_nature_2008}.  In the high wavelength
limit ($\lambda>2200$ nm), the $\mathrm{Re}(n)$ is zero and the
$\mathrm{Im}(n)$ is much larger than the $\mathrm{Re}(n)$, exhibiting metallic
behavior and transmission is equal to zero. This metallic behavior can be also
seen in the transmission, $T$, (see the supplementary material) for the many
layer structure. Above 2200nm, $T$ is low, and behaves as a metal, while for
$\lambda<2000$ nm, $T$ is relatively large ($\sim$ 0.8) and has Fabry-Perot
resonances structure. The $|\mathrm{Re}(n)$ shown in Fig. \ref{fig_n_converge}
converges between 1200 nm and 2200 nm to a finite value (positive for
wavelengths less than 1500 nm and negative for 1500 nm $<$ 2200 nm). For
$\lambda>$2200 nm, the $|\mathrm{Re}(n)|$ is zero and the $|\mathrm{Im}(n)|$ is
large of the order of 3, and as expected for large wavelengths this strongly
coupled metamaterial behaves as a metal. In addition, in Fig.
\ref{fig_n_converge} the 3-layer structure (the single fishnet structure) gives
results completely different than those for the strongly coupled fishnets.
These single fishnet results agree with those presented in Fig.
\ref{fig3_n_1_10_cell}. Another important quantity is the figure of merit (FOM)
which can be defined two different ways. The usual definition is
FOM=$|\mathrm{Re}(n)/\mathrm{Im}(n)|$ and the experimental definition of
$\mathrm{Im}(n)$ is given by $\mathrm{Im}(n)=(\lambda/4\pi d)\ln[(1-|R|)/|T|]$,
where $\lambda$, $d$, R, and T are the wavelength, sample thickness,
reflectance, and transmittance, respectively.\\

\begin{figure}[htb]
 \centering
  \includegraphics[width=8cm]{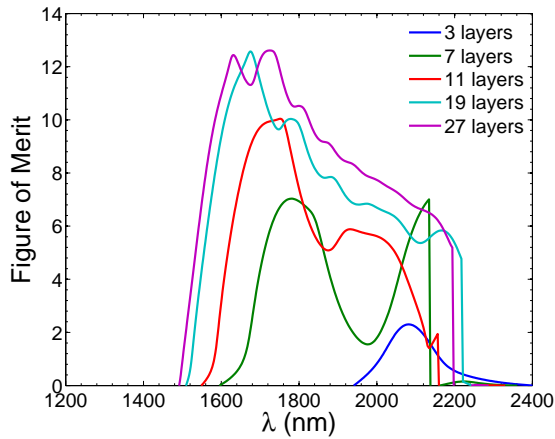}
  \caption{
    The figure of merit (FOM) of $\mathrm{Re}(n)<0$ region for 3, 7, 11, 19,
    and 27 layers strongly coupled fishnet structure.
    The FOM is calculated by FOM=$|\mathrm{Re}(n)/\mathrm{Im}(n)|$,
    where $\mathrm{Re}(n)$ is obtained by a retrieval procedure and
    $\mathrm{Im}(n)$ is calculated by
    $\mathrm{Im}(n)=(\lambda/4\pi d)\ln[(1-|R|)/|T|]$.
  }
  \label{fig_FOM}
\end{figure}

\section{Figure of merit calculations}
In Fig. \ref{fig_FOM} we present the results of the FOM as a function of
wavelength for different number of layers. For the one unit cell
fishnet (3 layers), the FOM is really small (of the order of 2) and
is located at $\lambda=$ 2100nm, the resonance frequency of the
single fishnet structure. As the number of layers increases, the FOM
increases and finally saturates to a constant value of the order of
10. This behavior of the FOM for the strongly coupled fishnets is
completely different for the weakly coupled fishnets, where the FOM
does not change dramatically \cite{zhou_photon_nano_2008} as one
uses more unit cells. Why is the FOM in the strongly coupled
fishnets so much different than the single fishnet? It has been
argued \cite{zhang_nature_2008, Li_Opex_2006} that the FOM is larger
because of the strong coupling between the neighboring layers, which
provides destructive interference of the antisymmetric currents
across the metal film and effectively cancels the current in the
center of the film, and, therefore, reduces the losses. We have
systematically studied the current density for the different number
of strongly coupled fishnet structures. For the single fishnet
structure, the current density is along opposite directions in the
two metallic bars. This is the typical behavior of negative index
materials. When the number of layers increases, the current density
is more complicated and there is no clear physical explanation why
one obtains negative $n$ and why the
FOM is so large.\\

\begin{figure}[htb]
 \centering
  \includegraphics[width=9cm]{fig6.eps}
  \caption{
    (a) The current density distribution for a 7 layers strongly coupled
    fishnet at wavelength, $\lambda=2230$ nm (antisymmetric mode), with
    $\mathrm{Re}(n)=-0.17$.
    (b) The current density distribution for a 7 layers strongly coupled
    fishnet at wavelength, $\lambda=1859$ nm (symmetric mode), with
    $\mathrm{Re}(n)=-2.5$.
    The cross-section is perpendicular to the y-axis
    (i.e., incident magnetic filed, {\bf H}, direction).
    The color shows the current density in
    x-direction, $J_x$, with the red and blue being the positive
    maximum and negative maximum of $J_x$, respectively. The arrows
    show the direction of current density inside the silver layers
    schematically.
  }
  \label{fig_current}
\end{figure}

In Fig. \ref{fig_current}, we present the current density along the
x-axis (or ${\bf E}$-direction as shown in Fig. \ref{fig_geom}),
$J_x$, of the antisymmetric and symmetric modes for the seven layers
(4 metallic layers and 3 dielectric layers) strongly coupled
structure. For the antisymmetric mode (as shown in Fig.
\ref{fig_current}(a)), two double-fishnets are formed by the first
and second silver layers, and by the third and fourth silver layers.
The induced current inside two double-fishnets excite the magnetic
fields, ${\bf B^\prime}$, along the same direction. However, the
second and the third silver layers also form a double-fishnet, which
excits the magnetic fields in the opposite direction. Therefore, the
excited magnetic fields, ${\bf B^\prime}$, are always anti-parallel
in the space between neighboring silver layers and cancel each
other. This explains the observation of a weak resonance with nearly
zero $n$. For the symmetric mode shown in Fig. \ref{fig_current}(b),
the first and the fourth silver layers have current density along
opposite directions and are almost uniform for all the metallic
thickness of 30 nm silver layers. In the second and third silver
layers the current density is no longer uniform in all thicknesses
of the silver layers. Instead, the current flows along opposite
directions on the two surfaces of each layer. Due to the
anti-parallel current on the surfaces of the second and third silver
layers, the induced magnetic field, ${\bf B^\prime}$, in the space
between neighboring silver layers, is always parallel to each other.
As a consequence, the 7 layers structure can be viewed as three
cascade double-fishnet structures with the induced magnetic fields,
${\bf B^\prime}$, along the same direction. Therefore, the symmetric
mode results in a strong resonance with large negative $n$. Our detailed
numerical work, shown in Fig. 4 and 6, explains very well why we
obtain very low $n\approx$0 at $\lambda$=2230 nm and why we obtain high
negative $n=$ -2.5 at $\lambda$=1859 nm. However, it is not clear that this
current density distribution is responsible for high FOM shown in Fig.
\ref{fig_FOM}.

\begin{figure}[htb]
 \centering
  \includegraphics[width=7cm]{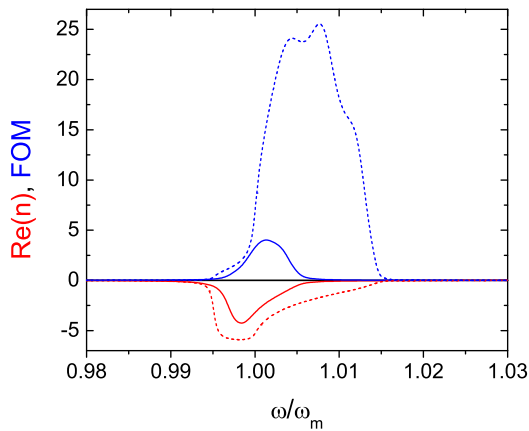}
  \caption{
    The real parts of refractive index (red), $\mathrm{Re}(n)$, and
    the figure of merit (blue), for the single layer fishnet structures
    with spacer thickness $s=0.025a$ (solid curves)
    and $0.1a$ (dashed curves), respectively.
    The other geometric parameters are given by $a_x=a_y=a$, $w_x=2a/5$,
    $w_y=a/3$, $t=a/300$, and the dielectric constant of the spacer
    is $\epsilon_r=5$.
  }
  \label{fig_periodicity_1layer}
\end{figure}

The reason that the single unit cell (metal-dielectric-metal) has low FOM or
high losses is due to its resonance structure. One way to increase the figure
of merit, which is the ratio of $|\mathrm{Re}(n)/ \mathrm{Im}(n)|$, is to move
away from the resonance frequency, where $\mathrm{Im}(n)$ is large, and
therefore FOM can increase dramatically. This can be accoplished in both the
weakly and strongly coupled fishnets, by introducing periodicity effects. For
the single unit cell fishnet, we can increase the size of the spacing layer and
one can see from Fig. \ref{fig_periodicity_1layer} that with a thicker spacing
layer, $s=0.1a$,  the $\mathrm{Re}(n)$ reaches the Brillouin zone and the
$\mathrm{Re}(n)<0$ region is extend into the area where
$\mathrm{Im}(n)\approx0$, so the FOM reaches a large value of 25..

\begin{figure}[htb]
 \centering
  \includegraphics[width=7cm]{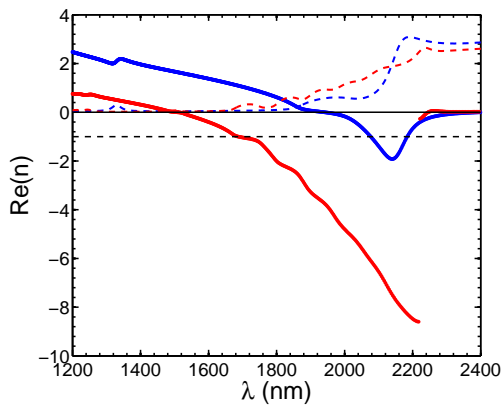}
  \caption{
    The real (solid curves) and imaginary (dashed curves)
    parts of refractive index, $\mathrm{Re}(n)$ and $\mathrm{Im}(n)$, for
    the 3 (blue) and 19 (red) layers fishnet structures. The black dash line
    shows the position where $\mathrm{Re}(n)=-1$.
  }
  \label{fig_periodicity}
\end{figure}

In Fig. \ref{fig_periodicity}, we present both the real and the
imaginary parts of the refractive index, $\mathrm{Re}(n)$ and
$\mathrm{Im}(n)$, for  the 3 and 19 layers fishnet structures. For
the 3 layers (the single layer of double-fishnet), the
$\mathrm{Re}(n)$ has a smooth resonance curve (blue solid). The
bandwidth of the $\mathrm{Re}(n)<0$ region is relative narrow and
close to the peak of $\mathrm{Im}(n)$ (blue dashed), so the figure
of merit is very small (as shown in Fig. \ref{fig_FOM}). For the
19-layer fishnets, the $\mathrm{Re}(n)$ curve (red solid) does not
have the resonance behavior expected for a single functional layer,
but it's very broad and has structure which is due to periodicity
effects \cite{PRB_koschny_retrieval_2005}. Notice that for the 19
layer structure $\mathrm{Re}(n)$=-1 at $\lambda=1688$ nm and the
$\mathrm{Im}(n)$ is 0.14, so the FOM is of the order of 10. However,
for the 3 layer structure, $\mathrm{Re}(n)$=-1 at $\lambda=2075$ nm
and 2185 nm, and the $\mathrm{Im}(n)$ is 0.44 and 1.43 respectively,
so the FOM is of the order of 1. Therefore, due to the distortion of
$\mathrm{Re}(n)$ caused by the periodicity effects, the FOM of the
fishnet structure increase dramatically as the number of layers
increases.

\section{Conclusions}
We have made a systematic study of the weakly and strongly coupled
fishnets to understand the origin of negative $n$, as well as the
origin of losses and the large value of the FOM for the strongly
coupled fishnets. We studied the size dependence of the retrieved
parameters ($\epsilon$, $\mu$, and $n$) of the weakly and strongly
coupled fishnet structures. For both cases we found the retrieved
parameters have a strong resonance behavior as the size of the unit
cell decreases. We have also studied the convergence of the
retrieved parameters, as the number of unit cells (layers) increase.
For the weakly coupled fishnet structures, we found the convergence
results are relatively close to the single unit cell. Also, the
converged FOM for the weakly coupled fishnet is the same order of
magnitude as the single fishnet. For the strongly coupled fishnet
structures, we demonstrated that hybridization happens and we have
two resonance modes. The antisymmetric resonance mode gives a strong negative
$n$. As more unit cells or layers are added, the convergence of the
retrieval parameters are completely different than the single
fishnet results and the FOM is much larger than the single fishnet.
We have demonstrated that the large FOM for the strongly coupled
fishnet is due to the periodicity effects.
%

%
\begin{acknowledgments}
We thank M. Wegener for useful discussions.
Work at Ames Laboratory was supported by the Department of Energy
(Basic Energy Sciences) under contract No. DE-AC02-07CH11358. This
work was partially supported by the Department of Navy, Office of
the Naval Research (Award No. N00014-07-1-0359), European Community
FET project PHOME (Contract No. 213390) and AFOSR under MURI grant
(FA 9550-06-1-0337).
\end{acknowledgments}

\appendix
\section{Effective parameter retrieval for strongly coupled fishnets}

\begin{figure}[htb]
 \centering
  \includegraphics[width=6.5cm]{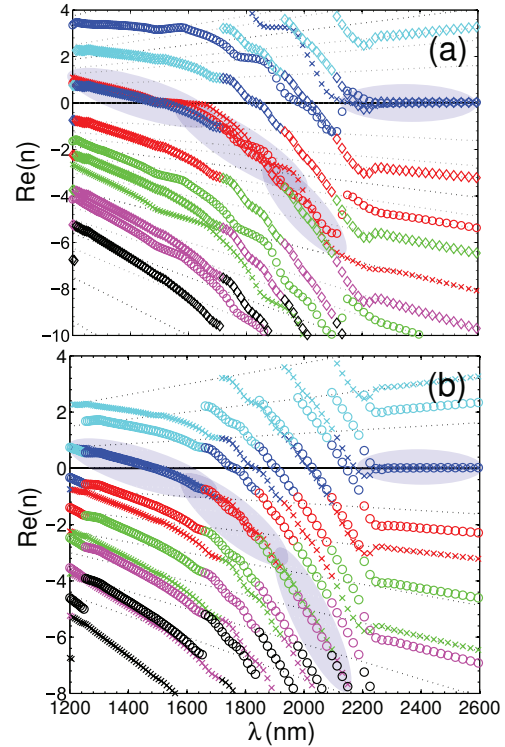}
  \caption{
    Branches of the refractive index, $\mathrm{Re}(n)$ with $m=1$ (cyan),
    $0$ (blue), $-1$ (red), $-2$ (green), $-3$ (magenta) and $-4$ (black).
    The cross, circle and diamond symbols in (a) represent
    $\mathrm{Re}(n)$ for 7, 11 and 19 layer strongly coupled fishnet
    structure, respectively.
    The cross and circular symbols in (b) represent the 19 and 27 layer
    strongly coupled fishnet structure, respectively.
    The shadow region shows where different branches overlap for 7, 11, 19,
    and 27 layers fishnet structures.
    The grey dotted lines show the branch boundaries which are
    given by $m\pi/kL$.
  }
  \label{fig_s1}
\end{figure}

The effective retrieved parameter ($\epsilon$, $\mu$, $n$ and $z$) of single
layer and many layers of metamaterial can be obtained from the transmission $T$
and reflection coefficient $R$. There is a need for $T$ and $R$ to be inverted.
As was discussed in detail in the literature [10-13], one can invert $T$ and $R$
\begin{eqnarray}\label{equ_z}
z(\omega)&=&\pm\sqrt{\frac{(1+R)^2-T^2}{(1-R)^2+T^2}}\\
n(\omega)&=&\pm\frac{1}{kL}\arccos\left(\frac{1-R^2+T^2}{2T}\right)+m\frac{2\pi}{kL}
\quad
\quad
\end{eqnarray}
where $L$ is the width of the homogeneous slab, and $m=\pm1$, $\pm2$, ... .
Note that both functions, $z(\omega)$ and $n(\omega)$, have multiple branches.
The correct branch for $z(\omega)$ is chosen by imposing the physical
requirement $\mathrm{Re}(z)\geq0$ which is due to causality. The problem with
the different branches of $\mathrm{Re}(n)$ can be solved by considering
different lengths for $L$, and one has to choose the branches that overlap.
Especially if one has many layers, then many branches exist and one has to be
very careful to select the correct ones. For the strongly coupled layers that
the results were presented in Fig. 4, we would like to discuss how these
branches were selected. The unit cell size is called $d_0$ and it consists from
metal-dielectric-metal and its width is $d_0=$160nm.

\begin{figure}[htb]
 \centering
  \includegraphics[width=6.5cm]{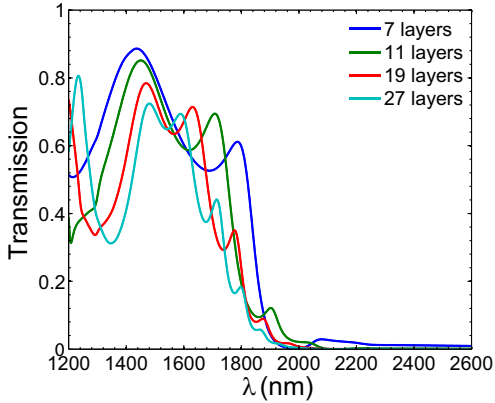}
  \caption{
    Transmission spectra for 7, 11, 19, and 27 layers strongly coupled fishnet system.
  }
  \label{fig_s2}
\end{figure}

In Fig.~9(a), we plot the branches and the retrieval results for 7 layers
(width$=2d_0$), 11 layers (width$=3d_0$) and 19 layers (width$=5d_0$). Notice
that the solutions for $\mathrm{Re}(n)$ overlap between 1200nm all the way to
2200nm and give negative values of $\mathrm{Re}(n)$. For $\lambda>2200$ nm,
$\mathrm{Re}(n)\approx0$ and converges and the $\mathrm{Im}(n)\approx3$ in this
region. So for $\lambda>2200$ nm the strongly coupled optical materials behave
as a metal. In Fig.~9(b), we plot the branches and the retrieved result for 19
layers (width$=5d_0$) and 27 layers (width$=7d_0$) and one can see clearly that
the convergence is much better for these larger systems. So we have solutions
consisting of two discontinued region for the $\mathrm{Re}(n)$,
$\mathrm{Re}(n)\approx0$ for $\lambda>2200$ nm and negative for 1500
nm$<\lambda<2200$ nm.

In Fig.~10, we present the results for transmission, $T$, versus wavelength.
Notice that for $\lambda>2200$ nm, $T \approx 0$, which is a metallic
behavior and this is the reason that $\mathrm{Re}(n)\approx0$ and
$\mathrm{Im}(n)\approx3$ for $\lambda>2200$ nm.

\end{document}